\documentclass[letterpaper, 10 pt, conference]{ieeeconf}  

\IEEEoverridecommandlockouts                              

\usepackage{amsmath} 
\usepackage{amssymb}  

\usepackage{tikz}
\usetikzlibrary{positioning}
\usetikzlibrary{shapes,arrows}
\usetikzlibrary{decorations.markings}

\usepackage{subfig}

\usepackage{comment}

\usepackage{amsthm}
\usepackage{cases} 

\newtheorem{theorem}{Theorem}

\newcommand{\R}{\ensuremath{\mathbb{R}}}
\newtheorem{rem}{\textnormal{\textbf{Remark}}}

\title{\LARGE \bf Nuclear fusion plasma fuelling with ice pellets using a \\
neuromorphic controller
}

\author{L.L.T.C. Jansen$^{1,2}$ \hspace{10 pt} E. Petri$^{2}$ \hspace{10 pt} M. van Berkel$^{1,3}$ \hspace{10 pt} W.P.M.H. Heemels$^2$%
\thanks{$^{1}$Loes Jansen and Matthijs van Berkel are
with DIFFER - Dutch Institute for Fundamental Energy Research, Eindhoven, The Netherlands.
        {\tt\small \{l.l.t.c.jansen, m.vanberkel\}@differ.nl}}%
\thanks{$^{2}$Loes Jansen, Elena Petri and Maurice Heemels are
with the Control Systems Technology (CST) Section of the Department of Mechanical Engineering, Eindhoven University
of Technology, The Netherlands.
        {\tt\small \{l.l.t.c.jansen, e.petri, m.heemels\}@tue.nl}}%
\thanks{$^{3}$Matthijs van Berkel is with the Department of Electrical Engineering, 
    Eindhoven University of Technology, The Netherlands.
        {\tt\small m.vanberkel@tue.nl}}%
\thanks{DIFFER is part of the institutes organisation of NWO. This work was supported by the research program VIDI with project number 19695, which is (partly) financed by NWO.}
\thanks{This work has been carried out within the framework of the EUROfusion Consortium, funded by the European Union via the Euratom Research and Training Programme (Grant Agreement No 101052200 - EUROfusion). Views and opinions expressed are however those of the author(s) only and do not necessarily reflect those of the European Union or the European Commission. Neither the European Union not the European Commission can be held responsible for them.
}
}

\begin{document}

\maketitle
\thispagestyle{empty}
\pagestyle{empty}

\begin{abstract}

In reactor-grade tokamaks, pellet injection is the best candidate for core plasma fuelling.
However, density control schemes that can handle the hybrid nature of this type of fuelling, i.e., the discrete impact of the pellets on the continuously evolving plasma density, are lacking.
This paper proposes a neuromorphic controller, inspired by the integrate-and-fire neuronal model, to address this problem.
The overall system is modelled as a hybrid system, and we analyse the proposed controller in closed loop with a single-input single-output linear time-invariant plasma model.
The controller generates spikes, representing pellet launches, when the neuron variable reaches a certain threshold. 
Between the control actions, or spikes, the system evolves in open loop.
We establish conditions on the controller variables and minimum actuator speed, depending on the reference value for the desired density, the pellet size and the time-constant of the plasma density,
that guarantee a practical stability property for the closed-loop system.
The results are illustrated in a numerical example.

\end{abstract}

\section{INTRODUCTION}

Nuclear fusion stands as one of the most promising future energy sources, with the potential to meet our growing energy demands while being virtually inexhaustible and inherently safe.
To achieve this, a tokamak is a device designed to harness fusion energy by creating and confining a plasma with strong magnetic fields. 
The fusion of deuterium and tritium within the plasma produces helium, neutrons, and energy, which is absorbed as heat by the vessel walls. 
This heat is then used to generate steam and electricity, similar to conventional power plants. 
To optimize energy output, it is crucial to control the plasma density by proper fuelling strategies. 
Higher particle density in the plasma leads to more fusion reactions and, consequently, more energy production.
On way of fuelling the plasma works by gas puffing, but this is only effective for low-density plasmas. 
It fails in high-performance tokamaks operating at higher densities, temperatures, and magnetic field strengths.
Under these conditions, gas particles ionise at the plasma edge, preventing deep penetration and making them ineffective for core fuelling \cite{Romanelli}.
An alternative fuelling method is based on pellet injection, which involves shooting frozen deuterium-tritium fuel pellets into the plasma. 
It is currently believed to be the best candidate for fuelling future high-performance machines \cite{Guelin-pellet}.

Controlling plasma density with pellet injection introduces challenges.
Ideally, the plasma density is maintained at a desired level by providing a continuously controlled particle flux to compensate for the ongoing losses.
However, since the pellet fired into the plasma, ablates and ionizes over a time of 1-2~ms \cite{Pegourie-ablation}, 
a duration that is negligible compared to typical control timescales,
it introduces particles in discrete bursts, causing sudden density spikes, and not the desired continuous particle flux.
This discrete nature must be considered in the control strategy, essentially leading to a hybrid systems control problem.

Although several experimental tokamaks are equipped with pellet injectors, there has been limited 
research on developing control schemes for these actuators.
In general, the (continuous) pellet fuelling rate is considered as control input \cite{Ravensbergen-control,pellets-JET,Lang-controller,Kudlacek-overview}, a straightforward choice that enables the use of standard linear controllers (as abstracted away from the discrete nature of pellets).
For instance, on the Axially Symmetric Divertor Experiment Upgrade (AUG), a control scheme for a centrifuge-type pellet actuator has been proposed and demonstrated \cite{Lang-controller}, \cite{Kudlacek-overview}. 
The control strategy uses the measured density in a PI feedback loop to output a particle flux.
This particle flux is then converted into a sequence of pellet injections with a separate sigma-delta modulator, as described in \cite{sigma-delta}.

Another control algorithm used at the Joint European Torus (JET) combines a simple PID controller with feedforward \cite{pellets-JET}. 
The desired particle input to the plasma was generated by either a continuous valve or discrete pellets, both considered continuous actuators using the same control loop with adjusted PID settings.
The authors of \cite{pellets-JET} state: ``With sufficient filtering the discrete nature of the pellets does not adversely affect the density control''.
However, while this works to some extent for current tokamaks, this approach is insufficient for density control using the larger fuelling pellets necessary for the operation of larger reactors such as ITER and DEMO, the successor of ITER, as was demonstrated in \cite{discrete-pellets1} and \cite{discrete-pellets2}.

As a result, for larger tokamaks, it is important to take the discrete nature of the pellets explicitly into account.
The model-predictive control (MPC) scheme proposed in \cite{Orrico-MPC1} is one of the first to treat pellets as discrete actuator decisions. 
Although this may work for larger tokamaks that allow longer times between two pellets, for currently available devices,
the implementation of this type of controller remains challenging in a real-time setting due to the significant computational effort required to solve the optimisation problems in MPC. 
Therefore, there is a strong interest for control solutions that are easier to implement in real-time while being effective in density control.

To address this need, in this paper, we introduce a hybrid control strategy based on a neuromorphic controller setup, see e.g. \cite{Petri}-\cite{Schmetterling}.
The controller we propose, is inspired by integrate-and-fire neuron models \cite{Lapicque,Spiking-neuron}, includes a first-order differential equation, describing the dynamics of the membrane potential of a neuron between spikes, and a mechanism to reset the membrane potential whenever a certain threshold is reached.
The similarities between the pellet fuelling effects on the plasma density, and a spike, or impulse input on a continuous plant, make this strategy a potentially suitable candidate for a pellet fuelling control scheme.
To be more precise, we use a variant of the neuromorphic controller proposed in \cite{Petri}, with adaptations to better mimic a pellet actuator.
A first adaptation is using a single neuron generating only positive impulses, as we can only add pellets to the plasma, and not remove any.
Second, the aim of this controller is to converge to a specified reference value rather than to the origin, as in \cite{Petri}.
Therefore, we evaluate the error between the desired reference value and the plasma density, instead of the plasma density alone.
Third, in \cite{Petri} spikes could occur at each moment of time, while for the pellet injection this can only occur at specific discrete moments that are periodic in time, synchronised with the actuator, as explained next.

There are two main types of pellet actuators, based on how the pellet gets its speed, necessary to travel deep into the plasma:
\begin{itemize}
    \item The \emph{centrifuge type actuator} drops a pellet on a spinning disc.
    The pellet gets its velocity from the centrifugal forces.
    Every time the pellet on the disc passes the launch opening, there is an opportunity to launch the pellet, so the controller needs to be synchronised to the centrifuge. 
    An example centrifuge can be found at AUG \cite{Ploeckl-centrifuge}, which can launch pellets at 
    a velocity of 560~m/s.
    \item In the \emph{gas gun actuator}, a pellet is launched by the pressure of the gas.
    Multiple stages can be used to build pressure, making it possible to reach higher pellet velocities.
    Whenever a pellet is ready to be launched, the controller can decide to do so whenever it desires.
    However, after a launch, the actuator needs to wait a certain preparation time to launch a next pellet.
    An example gas gun can be found at JET \cite{Combs-gasgun}, which can launch pellets with a velocity up to 1.4~km/s.
\end{itemize}
In this work, we will focus on the centrifuge pellet actuator, as it is easier to relieve the constraints afterwards to also fit the gas gun actuator.
In future work, we will consider the gas gun actuator and a corresponding neuromorphic controller, and compare it with the centrifuge.

The first main contribution of this work is the formulation of the structure of the neuron-inspired pellet controller. 
The second contribution is the formulation of a hybrid system model of the overall closed-loop system using the formalism of \cite{Goebel-hybrid}, where a jump corresponds to 
the firing of a pellet into the plasma.
The third contribution is that, for a linear time-invariant first-order plant, describing the plasma density dynamics, we prove a practical stability property of the closed-loop system, using a basic hybrid model
of the neuronal dynamics.
The formal analysis provides clear insights in the conditions on the parameters of the neuromorphic controller to obtain desirable behaviour.
Finally, we illustrate our new design in a numerical simulation.

\section{PRELIMINARIES}
\label{section:preliminaries}
The notation $\mathbb{R}$ stands for the set of real numbers, 
$\mathbb{R}_{\geq 0} := [0,+\infty)$ and
$\mathbb{R}_{> 0} := (0,+\infty)$.
We use $\mathbb{Z}$ to denote the set of integers, 
$\mathbb{Z}_{\geq 0} := \{0, 1, 2, \ldots \}$ and 
$\mathbb{Z}_{> 0} := \{1, 2, \ldots \}$.

We present the neuromorphic controller in closed loop with the plant, following the hybrid systems framework in \cite{Petri}, with the formalism in \cite{Goebel-hybrid}:
\begin{subnumcases}{\mathcal{H} : \label{eq:hybrid_system_general}}
    \hfill \dot{x} \; = \; F(x),  & $x \in  \mathcal{C}$ \label{eq:hybrid_system_general_flow}\\
    x^+ \; \in \; G(x),  & $x \in  \mathcal{D}$,\label{eq:hybrid_system_general_jump}
\end{subnumcases}
where $\mathcal{C} \subseteq \mathbb{R}^{n_x}$ is the flow set,
$\mathcal{D} \subseteq \mathbb{R}^{n_x}$ is the jump set,
$F : \mathbb{R}^{n_x} \to \mathbb{R}^{n_x}$ is the flow map, and 
$G : \mathbb{R}^{n_x} \rightrightarrows \mathbb{R}^{n_x}$ is the set-valued jump map,
with $n_x$ the dimension of the state vector $x$.
We consider hybrid time domains as defined in \cite[Definition 2.3]{Goebel-hybrid} and we use the notion of solutions for system (\ref{eq:hybrid_system_general}) as in \cite[Definition 2.6]{Goebel-hybrid}. 
Given a solution $x$ for system (\ref{eq:hybrid_system_general}) and its hybrid time domain dom $x$, we define 
$\sup_t  \text{dom } x:= \sup\{t \in \mathbb{R}_{\geq 0} : \exists \;j \in \mathbb{Z}_{\geq 0} \text{ such that } (t, j) \in \text{dom } x\} $ and
$\sup_j \text{dom } x:= \sup\{j \in \mathbb{Z}_{\geq 0} : \exists \; t \in \mathbb{R}_{\geq 0} \text{ such that } (t, j) \in \text{dom } x\}$. 
The notation $(t,j) \geq (t^\star, j^{\star})$ means that $t \geq t^\star$ and $j \geq  j^{\star}$,  where $(t,j),(t^\star,j^\star)\in\R_{\geq 0}\times\mathbb{Z}_{\geq0}$. Moreover, we say that the solution $x$ is complete if its hybrid time domain is unbounded and, in addition, is $t$-complete if $\sup_t \text{dom } x = +\infty$. 

\section{HYBRID MODEL FOR PLASMA DENSITY WITH PELLET INJECTION}
\label{section:hybrid_model}

To model our system in the hybrid framework of (\ref{eq:hybrid_system_general}), we separate the continuous evolution of the system between spiking control actions, called ``flow'', from the control actions, called ``jumps'', which correspond to pellets launches.
During flow, when the system is in open loop and no pellets are injected, 
we assume an exponential decay of the density, as a linear time-invariant first-order system, advocated in \cite{Derks} as a good approximation, i.e.,
\begin{equation}
    \dot{n}_e(t) = -\frac{1}{\tau}n_e(t)\text{,}
    \label{eq:openloop_density}
\end{equation}
where $n_e(t) \in \mathbb{R}_{\geq 0}$ is a scalar state, representing the electron density in the plasma at time $t \in \mathbb{R}_{\geq 0}$, and $\tau \in \mathbb{R}_{> 0}$ a time constant. 
We would like to regulate the density to a constant reference value $r \in \mathbb{R}_{\geq 0}$.
We define $x(t) := r - n_e(t) \in \mathbb{R}$ as the error between the desired reference value and the measured density.
During flows, using (\ref{eq:openloop_density}), $x(t)$ evolves as
\begin{equation}
    \dot{x}(t) = \frac{1}{\tau} (r - x(t))\text{.}
    \label{eq:openloop_x}
\end{equation}
Notice that this dynamics has some inherent boundedness property: since the electron density $n_e$ can never be negative, the error $x$ has the reference value $r$ as an upper bound.

\tikzstyle{block} = [draw, rectangle, 
    minimum height=3em, minimum width=2em]
\tikzstyle{sum} = [draw, circle, node distance=1cm]
\tikzstyle{input} = [coordinate]
\tikzstyle{output} = [coordinate]
\tikzstyle{pinstyle} = [pin edge={to-,thin,black}]
\tikzstyle{vec->}=[line width=1mm,-stealth,postaction={draw, line width=1mm, shorten >=4mm, -}]
\tikzset{->-/.style={decoration={
  markings,
  mark=at position #1 with {\arrow{>}}},postaction={decorate}}}

\begin{figure}
    \centering
    \begin{tikzpicture}[auto, node distance=4em, scale=0.8, every node/.style={scale=0.8},
        sat/.style = {draw, box=#1,
                     path picture={\draw[very thick] 
                            (\ppbb.center) -- ++ (-1mm,-3mm) -- ++ (-2mm,0)
                            (\ppbb.center) -- ++ (+1mm,+3mm)
                            ;}
                     },
                     >=latex']
    \node [input](input){$e$};
    \node [block, right of=input, minimum width=2em, node distance=7em](neuron){\begin{tabular}{c} Controller \\ $\xi$,T \\$\Delta$,$\alpha$,$T_c$ \end{tabular}};
    \node [block, right of=neuron, minimum width=3em, node distance=8em](plant){plant};
    \node [sum, right of=plant, node distance=6em](sum){$\Sigma$};
    \node [input, above of=sum, node distance=4em](ref){ref};
    \node [input, right of=sum, node distance=6em](tmp1){};
    \node [input, left of=sum, node distance=2em](min){};
    \node [input, above of=sum, node distance=2em](plus){};
    \node [output, minimum width=2em, node distance=2em, right of=tmp1](output){$e$};
    \node [output, minimum width=2em, node distance=4em, below of=tmp1](tmp2){};

    \node [input, right of=neuron, above of=neuron, node distance=2.6em](spikea){};
    \node [input, right of=spikea, node distance=3em](spikeb){};
    \node [input, right of=spikea, node distance=0.4em](spike11){};
    \node [input, above of=spike11, node distance=1em](spike12){};
    \node [input, right of=spikea, node distance=1.5em](spike21){};
    \node [input, above of=spike21, node distance=1em](spike22){};
    \node [input, right of=spikea, node distance=2em](spike31){};
    \node [input, above of=spike31, node distance=1em](spike32){};
    \node [input, right of=spikea, node distance=2.8em](spike41){};
    \node [input, above of=spike41, node distance=1em](spike42){};
    \draw [-]  (spikea) -- (spikeb);
    \draw [-]  (spike11) -- (spike12);
    \draw [-]  (spike21) -- (spike22);
    \draw [-]  (spike31) -- (spike32);
    \draw [-]  (spike41) -- (spike42);

    \draw [draw,->] node [near end] {} (input) -- (neuron);
    \draw [->] (neuron) -- node {$u$} (plant);
    \draw [-] (plant) -- node{$n_e$} (min);
    \draw [->] (min) -- node{$-$} (sum);
    \draw [->] (sum) -- node {$x=r-n_e$} (output);
    \draw [-] (ref) -| node {$r$} (plus);
    \draw [->] (plus) -| node {$+$} (sum);
    \draw [-] (tmp1) -- (tmp2) -| (input);
    \end{tikzpicture}
  \caption{System with neuromorphic controller, reference input $r$, plasma density $n_e$, error $x$ and control action $u$ consisting of Dirac pulse (spike) trains. 
  The states of the controller are the neuron membrane potential $\xi$ and timer $T$.
  Moreover, the controller parameters are the threshold $\Delta$, the effect of one pellet on the plasma $\alpha$, and the time between launch slots $T_c$.
  Only positive values of $x$ are used as input to the controller. 
}
\label{fig:block_diagram}
\end{figure}
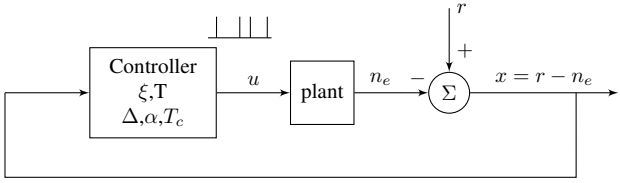

In our control configuration, the error $x$ is the input to the controller.
The block diagram, representing the closed-loop system architecture, is shown in Fig.~\ref{fig:block_diagram}.
The controller generates pellet launches, which will be considered as spike inputs to the plant.
While (\ref{eq:openloop_x}) describes the continuous, open-loop dynamics between pellet launches, we model the effect of the controller spike as a jump in (\ref{eq:hybrid_system_general}).
A spike can be considered as a Dirac delta pulse $m_p\delta(t-t_j)$, with $m_p \in \mathbb{R}_{>0}$ the spike amplitude, representing the size of a pellet, 
time $t \in \mathbb{R}_{\geq 0}$ and spiking time $t_j$, with $j \in \mathbb{Z}_{>0}$.
A spike, or pellet launch, will increase the density $n_e$ and thus will decrease the error $x$ during jumps, which becomes, in the hybrid setting of (\ref{eq:hybrid_system_general_jump})
\begin{align}
    x^+ & = x - \alpha\text{,}
    \label{eq:jump_x}
\end{align}
where $\alpha := B m_p$, and $B \in \mathbb{R}_{>0}$ is the conversion from pellet size (number of particles) to its effect on the density (particles per m$^3$).
The objective is now to design a controller, see Fig. \ref{fig:block_diagram}, generating the spikes, or pellet launches, such that a practical stability is satisfied for (\ref{eq:openloop_x}), (\ref{eq:jump_x}) in the sense that $x$ converges asymptotically to a (small) neighbourhood of 0.

\section{NEUROMORPHIC CONTROLLER}
In this section, we propose a neuromorphic controller, inspired by a spiking neuron, taking the error $x$ as an input.
The neuron has an accumulator $\xi \in \mathbb{R}_{\geq 0}$ and 
it generates pellet launches whenever $\xi$ reaches a certain threshold, denoted by $\Delta \in \mathbb{R}_{>0}$, after which it resets to 0.
Pellets can only be launched when the actuator is ready \cite{Ploeckl-centrifuge}, necessitating synchronisation of the trigger with the actuator's availability. 
This is achieved by checking the threshold condition only at times when a pellet launch slot is available.
The parameter $\xi$ can be considered as the neuron membrane potential and its dynamics are inspired by the integrate-and-fire neuron model of \cite{Lapicque,Spiking-neuron}.
Hence, during flow the dynamics of $\dot \xi$ are
\begin{equation}
    \dot{\xi} = \text{max}(0,x) = 
    \begin{cases}
        0  & \text{if} \quad    x \leq 0 \\
        x & \text{if} \quad     x > 0 \text{.}
    \end{cases}
    \label{eq:neuron_flow}
\end{equation}
Negative values of $x$ are thus ignored in the dynamics, and the membrane potential $\xi$ cannot decrease during flow.
This prevents the controller from acting when the density is already above the reference value.
The neuron triggers a spike when
\begin{equation}
    \xi \geqslant \Delta
    \label{eq:jump_cond_neuron}
\end{equation}
is satisfied,
and a pellet launch slot is available.
The latter condition is formulated by adding a timer variable $T$ with dynamics
\begin{equation}
    \dot{T} = 1 \text{,}
    \label{eq:timer_flow}
\end{equation}
where $T \in \mathbb{R}_{\geq 0}$.
Every time a launch slot is available, i.e., when
\begin{equation}
    T \geqslant T_c \text{,}
    \label{eq:jump_cond_timer}
\end{equation}
with $T_c \in \mathbb{R}_{> 0}$ the time between launch slots (in a centrifugal type launcher \cite{Ploeckl-centrifuge}),
the controller has to decide to launch a pellet or skip that available slot. In both cases, the timer is reset, as 
\begin{equation}
    T^+ = 0 \text{.}
    \label{eq:reset_timer}
\end{equation}
The neuron membrane potential $\xi$ is only reset, if a pellet is fired, giving 
\begin{equation}
    \xi^+ = 0 \textcolor{blue}{,}
    \label{eq:reset_neuron}
\end{equation}
If the timer is reset due to an available launch slot but no pellet is fired, $\xi$ is not modified, i.e., $\xi^+ = \xi$.

To capture all of the above, we define the overall state as $q := (x,\xi,T) \in \mathcal{X}$ with
$\mathcal{X}:=\{(x,\xi,T) \in \mathbb{R}^3 : (x \leq r) \wedge (\xi \geq 0) \wedge (T \geq 0) \}$ and we obtain the hybrid system
\begin{equation}
    \begin{cases}
        \hfill  \dot{q} \; = \; F(q), & q \in  \mathcal{C} \\
        q^+  \in \;  G(q),  & q \in  \mathcal{D} \text{, }
    \end{cases}
    \label{eq:hybrid_system}
\end{equation}
where the flow map $F$ is defined, for any $q \in \mathcal{C}$, from (\ref{eq:openloop_x}), (\ref{eq:neuron_flow}) and (\ref{eq:timer_flow}),
\begin{equation}
    F(q) := 
    \begin{pmatrix}
        -\frac{1}{\tau}x + \frac{r}{\tau} \\
        \max(0,x) \\
        1
    \end{pmatrix}\text{.}
    \label{eq:flowmap}
\end{equation}
The flow set $\mathcal{C}$ in (\ref{eq:hybrid_system}) is defined as
\begin{equation}
    \mathcal{C} := \{ q \in \mathcal{X}: (T \leq T_c) \}\text{.}
    \label{eq:flowSet}
\end{equation}

The jump set $\mathcal{D}$ in (\ref{eq:hybrid_system}) is defined as, from (\ref{eq:jump_cond_timer}),
\begin{equation}
    \mathcal{D} := \{ q \in \mathcal{X} : T \geqslant T_c \}. \label{eq:jumpset}
\end{equation}
Note that the way $\mathcal{C}$ and $\mathcal{D}$ are defined jumps take place when $T>T_c$, i.e. (\ref{eq:jump_cond_timer}).
To write the jump map, we write $\mathcal{D}$ as 
$\mathcal{D} := \mathcal{D}_1 \cup \mathcal{D}_2$
where
\begin{equation}
    \mathcal{D}_1 := \{ q \in \mathcal{X}: (T \geq T_c ) \wedge (\xi \leq \Delta )\} \label{eq:jumpsetD1}
\end{equation}
and
\begin{equation}
    \mathcal{D}_2 := \{ q \in \mathcal{X}: (T \geq T_c ) \wedge ( \xi \geq \Delta )\}\text{.}
    \label{eq:jumpsetD2}
\end{equation}

The jump map $G$ in (\ref{eq:hybrid_system}) is defined for any $q \in \mathcal{D}$, from (\ref{eq:jump_x}), (\ref{eq:reset_timer}) and (\ref{eq:reset_neuron}), as
\begin{equation}
    G(q) := G_1(q) \cup G_2(q)
    \label{eq:jump_map}
\end{equation}
with
\begin{equation}
    G_{1}(q) := 
    \begin{cases}
    \left\{
        \begin{pmatrix}
            x \\
            \xi \\
            0
        \end{pmatrix} 
        \right\}
        & q \in \mathcal{D}_1 \\
        \hfill \emptyset \hfill  & q \notin \mathcal{D}_1,
    \end{cases}
    \label{eq:jump_map_G1}
\end{equation}
corresponding to no pellet launch, and
\begin{equation}
    G_{2}(q) := 
    \begin{cases}
    \left\{
        \begin{pmatrix}
            x - \alpha \\
            0 \\
            0
        \end{pmatrix} 
        \right\}
        & q \in \mathcal{D}_2 \\
        \hfill \emptyset \hfill  & q \notin \mathcal{D}_2
    \end{cases}
    \label{eq:jump_map_G2}
\end{equation}
related to a pellet launch.

\begin{rem}\label{rem:completeness}
    The hybrid model in \eqref{eq:hybrid_system}-\eqref{eq:jump_map_G2} satisfies the hybrid basic conditions \cite[Assumption 6.5]{Goebel-hybrid}, which guarantee useful well-posedness properties. Moreover, by using \cite[Proposition~6.10]{Goebel-hybrid} and careful tangent cone computations, it is not hard to show that non-trivial solutions exist for all initial states in $\mathcal{C}\cup \mathcal{D}$. Due to the fact that finite escape times can be ruled out during flow (linear growth bounds on dynamics) and $G(\mathcal{D}) \subset \mathcal{C}$, all maximal solutions are complete, i.e., have unbounded hybrid time-domains. Using the particular structure of the hybrid model, due to the presence of the timer $T$, which reminds of periodic event-triggered control setups, see, e.g., \cite{WanPos_TAC20a, HeeDon_TAC13a}, we can guarantee that Zeno behaviour can be excluded, as we have a dwell-time of at least $T_c$ time units between jumps. Hence, maximal solutions are $t$-complete, i.e., $\sup_t  \textnormal{dom } q = +\infty$. 
\end{rem}

Now that we have proposed the neuromorphic controller to control the plasma density and an appropriate hybrid closed-loop system model, it is our objective to derive conditions on the neuron parameters $\alpha$ and $\Delta$, and the actuator parameter $T_c$, that guarantee a practical stability property of system (\ref{eq:hybrid_system}).

\section{MAIN ANALYSIS}
In this section we present our main analysis result. 
\begin{theorem}
\label{theorem1}
Consider system (\ref{eq:hybrid_system}), 
with $F, \mathcal{C}, G, \mathcal{D}$ as in \eqref{eq:flowmap}-\eqref{eq:jump_map_G2},
with a desired reference value $r \in \mathbb{R}_{>0}$, jump size $\alpha \in \mathbb{R}_{>0}$, and $r>\alpha$.
For any 
\begin{equation}
    T_c \in \left(0,\tau \ln{\frac{r}{r-\alpha}}\right]\text{,}    
    \label{eq:Tc_condition}
\end{equation}
select 
\begin{equation}
\Delta \in \left[0,r \tau \ln{\frac{r}{r-\alpha}} - r \tau \left(1- \frac{r-\alpha}{r} e^{\frac{T_c}{\tau}}\right) - r T_c\right]\text{}
\label{eq:Delta_condition}
\end{equation}
and define $\tau_d := \tau\ln{(\frac{r-\alpha}{r})} \in \R_{>0}$. 
Then, any solution $q$ of the hybrid system (\ref{eq:hybrid_system}) with $\xi(0,0)=0$, satisfies for all $(t,j) \in \textnormal{dom }q$, 
\begin{equation}
\begin{aligned}
 -\alpha < x(t,j) \leq \gamma^{\left(\frac{t}{\tau_d} -1\right)}  x(0,0) + \alpha \text{,} &\\
   \text{if } \;  x(0,0) &>0 \\
   \min\left(r-e^{-\frac{t}{\tau}}\left(r-x(0,0)\right),-\alpha\right) < x(t,j) \leq  \alpha \text{,} &\\
   \text{if } \;  x(0,0) &\leq 0 \text{,}
\end{aligned}
\label{eq:ultimate_bound}
\end{equation}
where $\gamma := \frac{r-\alpha}{r} \in (0,1)$.
Hence, for all maximal solutions we have $\lim_{t+j \rightarrow \infty} \sup |x(t,j)| \leq \alpha$.
\end{theorem}

Theorem~\ref{theorem1}, whose proof is given in the Appendix, 
provides conditions on the controller parameters $\alpha$, $\Delta$, $T_c$, to ensure that the proposed neuromorphic controller guarantees a practical stability property for the system (\ref{eq:hybrid_system}). 
When the conditions (\ref{eq:Tc_condition}), (\ref{eq:Delta_condition}) are met, 
the density error $x$ converges asymptotically to a neighbourhood $(-\alpha,\alpha]$ of the origin. 
Indeed, due to the $t$-completeness property (see Remark~\ref{rem:completeness}), \eqref{eq:ultimate_bound} ensures that $\lim_{t+j \rightarrow \infty} \sup |x(t,j)| \leq \alpha$. 

The condition on $T_c$ in (\ref{eq:Tc_condition}) can be interpreted as an upper limit on the minimum time between two pellet launches, or, in general, how fast the actuator must be.
This condition is necessary to realise the practical stability property.

A few other comments are in order regarding the theorem.
The condition that the reference input $r$ is larger than the effect of the pellet on the plasma density $\alpha$,
is a natural condition, as in the case where $r \leq \alpha$, control is actually not helpful due to the discrete nature of the actuator.
Indeed, system (\ref{eq:openloop_x}), which is asymptotically stable with $x$ converging to $r$ over time, has a better stability guarantee than Theorem~\ref{theorem1} in that case, so
a reference $r \leq \alpha$ is not meaningful.
Note that the ratio between $r$ and $\alpha$ is important for the actuator speed,
as we can rewrite (\ref{eq:Tc_condition}) as
\begin{equation}
    T_c \in \left(0,\tau \ln{\frac{\frac{r}{\alpha}}{\frac{r}{\alpha}-1}}\right]\text{.}    
\end{equation}
A large reference input $r$, compared to $\alpha$, lowers the upper limit of $T_c$, meaning it requires a faster actuator.
However, there is usually a physical limit on $T_c$, imposed by the actuator.
In practice, given a $T_c$ and $\alpha$ (related to the pellet size), 
we may have a limit on the maximum reference input that can be achieved with this actuator.
If we want a higher reference value with the given timing constraints on the actuator, a solution can be to increase the size of a pellet, leading to an increase of $\alpha$, the effect the pellet has on the plasma density.

The neuron threshold $\Delta$ can be interpreted as a tuning knob for the controller
and should be in the bounds given by (\ref{eq:Delta_condition}).
Depending on the value chosen for $\Delta$, this controller can react fast to an occurring error ($\Delta \approx 0$), or the controller can be ``slow'' ($\Delta \approx$ the upper bound in (\ref{eq:Delta_condition})). 
This tuning effect will be demonstrated in Section \ref{section:numerical_sim}.
At first glance, the usefulness of tuning $\Delta$ may appear limited, after all we can choose $\Delta=0$, leading to the simple control scheme of shooting a pellet at the next available timeslot whenever an error occurs.
However, establishing precise bounds on $\Delta$ proves valuable in several key scenarios. 
For instance, realistic operating conditions introduce additional complexity: the particle confinement time $\tau$ varies across operating regimes, actuator parameters $\alpha$ and $T_c$ experience disturbances, and density measurements $n_e$ contain noise.
By first defining allowable operating range for $\Delta$, this work lays the foundation for future robustness analysis against such perturbations.

The neuron-like variable $\xi$ is initialized at $\xi(0,0)=0$. This is a design choice to simplify the proof of Theorem~\ref{theorem1}, but any other choice for the initial condition would not compromise the stability result.
As the system is inherently stable, a simple translation in the hybrid time $(t,j)$, making the first jump in $\xi$ the new starting point, $(t_1,1) \rightarrow (0,0)$, will result in a system in which the conditions, and thus the result, from Theorem~\ref{theorem1} still holds.

\section{NUMERICAL SIMULATION}
\label{section:numerical_sim}
We consider a tokamak with a volume of around 13~m$^3$ with pellets consisting of $1.3 \times 10^{20}$~particles. 
A pellet increases the plasma electron density by $\alpha =1 \times 10^{19}$ particles/m$^3$.
The desired reference value $r = 7 \times 10^{19}$~particles/m$^3$.
The pellet actuator is a centrifuge working at 100~Hz, resulting in $T_c=0.01$~s.

\begin{figure}[ht!]
     \centering
     \subfloat[][Fast controller, $\Delta \approx 0$: reacts as soon as an error occurs, pushing the lower limit to the minimum of the bound.]{\includegraphics[width=0.9\linewidth]{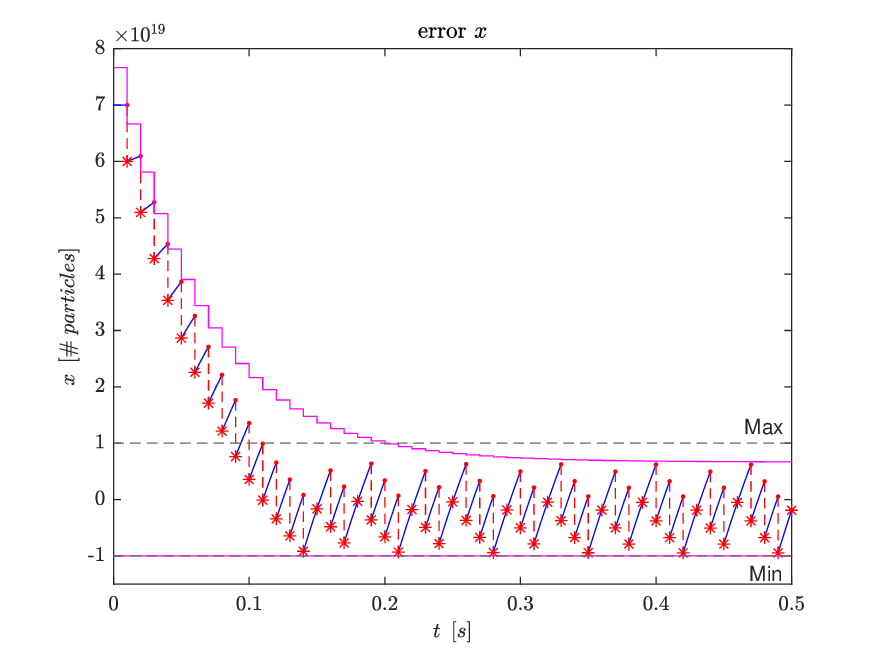}}
     \\
     \subfloat[][Slow controller, $\Delta = 10^{16}$: reacts later to an error, letting the error hit the maximum of the bound.]{\includegraphics[width=0.9\linewidth]{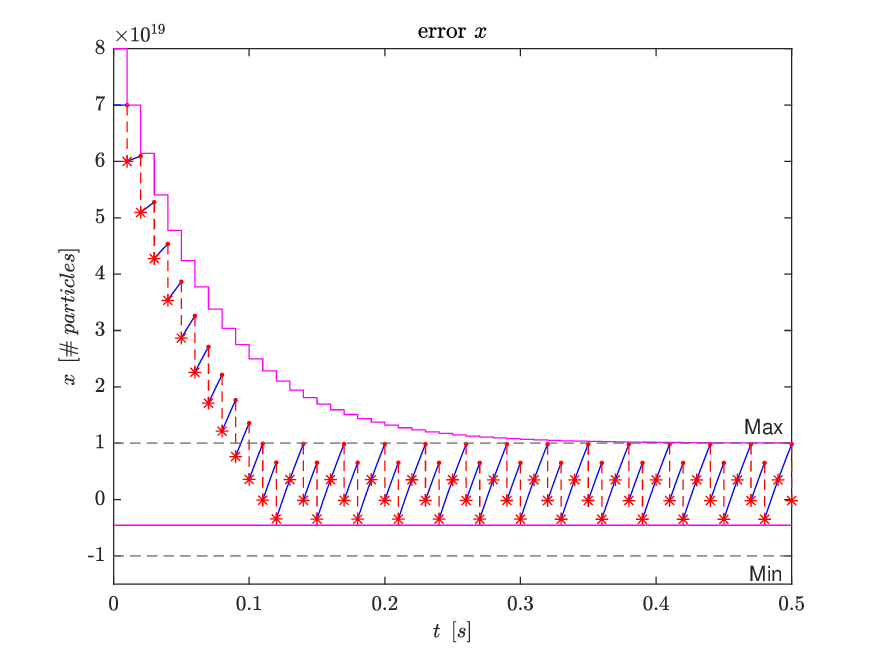}}
     \caption{Comparison of fast and slow controller by tuning neuron threshold $\Delta$. The blue line denotes the density error $x$. 
     The magenta dashed lines are the bounds established in Theorem~\ref{theorem1}.}
     \label{fig:fast_vs_slow_controller}
\end{figure}

We now apply the neuromorphic controller to the system described in Section \ref{section:hybrid_model} with $\tau = 0.1$~s.
From condition (\ref{eq:Tc_condition}) of Theorem \ref{theorem1}, we get $T_c \in (0, 0.0154]$.
With $T_c = 0.01$ in the interval, we can choose $\Delta \in (0,1.0080 \times 10^{16}]$.
We consider two possible options for $\Delta$ in the following. 
In particular, we first consider  $\Delta =1$, which is the fastest controller we can have, and then we compare it to the slower controller, with $\Delta = 10^{16}$.
The results obtained are shown in Fig.\ref{fig:fast_vs_slow_controller}.

The blue line indicates the continuous flow of the error, a red~$\ast$ shows when the actuator is ready to launch a pellet, the red dashed lines show the jumps in the density when a pellet is actually fired.
Fig. \ref{fig:fast_vs_slow_controller} confirms that the proposed neuromorphic controller satisfies a practical stability property of the closed-loop system, as proven in Theorem \ref{theorem1}.
The minimum and maximum value of the ultimate bound are shown in black dashed lines at -10$^{19}$ and 10$^{19}$ particles/m$^3$.
The transient effect on the bound, given in (\ref{eq:ultimate_bound}), is shown in magenta.
With the timing constraint on $T_c$ met, we can use the neuron parameter $\Delta$ to tune the speed of the controller and influence where, within the region of the ultimate bound, the system converges to.

In addition, we can see both controllers use the maximum amount of pellet slots available in the beginning, when the error is high. 
The slow controller is the first to slow down and use fewer launch slots to fire pellets.
For both controllers a steady state behaviour is reached that stays within the $\pm \alpha$-bounds as obtained from Theorem \ref{theorem1}.

Note that when the centrifuge timing $T_c$ gets closer to the upper bound imposed by Theorem \ref{theorem1},  the range in which we can choose $\Delta$ becomes more limited, i.e., a slower actuator requires a faster controller and we lose the margin for tuning the controller.
However, the region the system converges to, will always be in between the bounds given by Theorem~\ref{theorem1}.

\section{CONCLUSIONS}
We proposed a neuromorphic controller to control the plasma electron density in a tokamak using discrete pellet fuelling.
The controller is based on a neuron variable, whose dynamics are inspired by the integrate-and-fire neuron model.
We modelled this new neuromorphic controller in closed loop with a single-input single-output linear system capturing the plasma density dynamics in a hybrid systems framework.
Using this hybrid system model, we have provided a formal analysis, demonstrating that the neuromorphic controller guarantees a practical stability property for the closed-loop system,
if the neuron parameters are chosen according to the derived conditions.
Finally, a numerical example illustrated our analysis and design.
Note that the described model is intentionally simple, and although different methods could be suitable, the neuromorphic controller was chosen as it matches the actuators' spiking nature.

There are many possible avenues for future work in this area, that can build on the basic results presented in this paper.
\paragraph{Comparison to existing controllers}
Analysing the present AUG controller, where an integrate-and-fire scheme with a fixed $\Delta$ is used in combination with PI loop gains.
\paragraph{Use of gas gun actuator} 
This actuator does not need synchronisation but it has a waiting time.
An important adaptation of the neuron model is implementing a refractory period, a time the neuron has to wait after generating a spike, which can be linked to the pellet preparation time needed in both a centrifuge and a gas gun actuator.
\paragraph{Ramp up/down}
Expanding the analysis of the reference input from a fixed set-point to a ramp up or ramp down scenario, which, if successful, would make the controller suitable for use during startup or shutdown of a tokamak as well.
\paragraph{Time delay}
The way the actuators are built, with long guiding tubes between the launcher and the plasma, causes a time delay between firing a pellet and its arrival in the plasma.
In some devices, this delay is significant, and it needs to be taken into consideration, e.g., at AUG it takes 80~ms between pellet launch and its arrival in the plasma, while the centrifuge can shoot a pellet every 14~ms.
\paragraph{Packet loss}
We would like to include disturbances in the system and assure that the controller is robust against pellet loss, as in practice, part of a pellet, or sometimes even an entire pellet, can get lost in the guiding tubes. 
For instance in ITER, up to 15\% of pellets are expected to be lost, and even when successfully injected, sublimation can cause around 10\% uncertainty in pellet mass.
\paragraph{Disturbances on dynamics}
An example of a disturbance can be jitter, either in the variation of the centrifuge speed, causing variations in the $T_c$ we discussed here, or in the variation of the time delay between the firing and the density increase.
\paragraph{All of the above integrated/combined}
The combination of the pellet travel time, the disturbances, and the losses in the system, make this a control challenge far from trivial, and studying these effects in combination is left as future work.
\paragraph{More complex plasma models}
Instead of the basic first-order plasma model with a fixed time constant used here, we could simulate the controller on a more complete, high-fidelity model of a tokamak, such as JINTRAC, or even extend the formal analysis.
\paragraph{Testing}
Finally, it would be of interest to run the proposed neuromorphic controller on a real tokamak.

\appendix
\noindent\textbf{Proof of Theorem \ref{theorem1}.}
Let $q$ be a maximal solution to system (\ref{eq:hybrid_system}), and thus 
$x(0,0) \leq r$. Note that $q$ is $t$-complete (see Remark \ref{rem:completeness}).
Pick any $(t,j) \in \text{dom } q$ and let 
$0 = t_0 \leq t_1 \leq \ldots \leq t_{j+1} = t$
satisfy $\text{dom }q \ \cap \ ([0,t] \times \{0,1,\ldots,j\}) = \bigcup_{i=0}^j [t_i,t_{i+1}] \times \{i\}$.
Note that $t_{j+1}=t$ is not necessarily a jump time but $t_1,\ldots,t_j$ are jump times.
To separate the jumps where a pellet is being fired, from the jumps where no pellet is fired, 
we define the set of hybrid times at which a jump occurs 
due to a spike generated by jump set $\mathcal{D}_2$, i.e., due to a pellet being fired, as
\begin{multline}
    \mathcal{P}(q) := \{(t, i) \in \text{dom } q : q(t,i) \in \mathcal{D}_2 \text{ and } \\
q(t, i + 1) \in G_2(q(t, i)) \}.
\end{multline}

\noindent \textbf{Some general expressions of solutions}

For all $i \in \{0,1,\ldots,j\}$ and all $s \in [t_i,t_{i+1}]$,
$q(s,i) \in \mathcal{C}$  
and from (\ref{eq:openloop_x}), we have 
\begin{equation}
    x(s,i) = r-e^{-\frac{1}{\tau}(s-t_i)}\left(r-x(t_i,i)\right).
    \label{eq:x_s_i}
\end{equation}

From (\ref{eq:x_s_i}) we get the increase in $x$ between two jumps, for each $i \in \{0,1,\ldots,j\}$,
\begin{equation}
    x(t_{i+1},i) = r-e^{-\frac{1}{\tau}(t_{i+1}-t_i)}\left(r-x(t_i,i)\right).
    \label{eq:V_increase_between_jumps}
\end{equation}

Similarly, for each 
$i \in \{0,1,\ldots,j\}$ and all $s \in [t_i,t_{i+1}]$,
from the neuron dynamics (\ref{eq:neuron_flow}), we have
\begin{equation}
    \xi(s,i) = \xi(t_i,i) + \int_{t_i}^{s} \text{max}\left(0,x(\tilde{s},i)\right)d\tilde{s}.
\end{equation}
Consequently, when $x(t_i,i) > 0$, which we assume for now (later we consider $x(t_i,i) \leq 0$), for all $s \in [t_i,t_i+1]$,
\begin{equation}
    \xi(s,i) = \xi(t_i,i) + \int_{t_i}^{s} x(\tilde{s},i) d\tilde{s} .
    \label{eq:xi_s_i_zonder_max}
\end{equation}

For each 
$i \in \{1,\ldots,j\}$,
$q(t_i,i) \in \mathcal{D}$,
and using the jump map in (\ref{eq:jump_map})-(\ref{eq:jump_map_G2}), we have that
\begin{equation}
    \begin{aligned}
        \xi(t_i,i) &= \xi(t_i,i-1) &\text{ when } (t_i,i-1) \notin \mathcal{P}(q) \\
        \hfill \xi(t_i,i) &= 0 \hfill &\text{ when } (t_i,i-1) \in \mathcal{P}(q) .
    \end{aligned}
\end{equation}
\noindent \textbf{Increase of $x$ during flow between two pellets launches}

Let $(t_i,i-1) \in \mathcal{P}(q)$ and, using jump set $\mathcal{D}_2$ from (\ref{eq:jumpsetD2}), define
\begin{multline}
    p(i) := \text{inf}\{ \tilde{i} \geq i : \exists s \in \mathbb{R}_{\geq 0} \\
    \text{ such that } (s,\tilde{i}) \in \mathcal{P}(q) \}.
    \label{eq:pellet_jump_time}
\end{multline}
Note that $t_{p(i)}$ is the first time after $t_i$ at which a pellet is fired,
and that, between $t_i$ and $t_{p(i)}$ only jumps generated by the timer occur, where no pellet is fired.
Consider that there are $n \in \mathbb{Z}_{\geq 0}$ of these jumps.
From (\ref{eq:xi_s_i_zonder_max}) we have that for all $m \in \{0,1,\ldots, n-1\}$ and all $s \in [t_{i+m},t_{i+m+1}]$,
\begin{equation}
    \xi(s,i+m) = \xi(t_{i+m},i+m) + \int_{t_{i+m}}^{s} x(\tilde{s},i+m) d\tilde{s} .
    \label{eq:xi_flow_tussenin}
\end{equation}
Similarly, for all $s \in [t_{i+n},t_{p(i)}]$,
\begin{equation}
    \xi(s,i+n) = \xi(t_{i+n},i+n) + \int_{t_{i+n}}^{s} x(\tilde{s},i+n) d\tilde{s} .
    \label{eq:xi_flow_voor_pellet_spike}
\end{equation}
Moreover, recalling that when $(t_k,k-1) \notin \mathcal{P}(q)$, $\xi(t_k,k) = \xi(t_k,k-1)$, from (\ref{eq:xi_flow_tussenin}) and (\ref{eq:xi_flow_voor_pellet_spike}) we obtain, for all $s \in [t_i,t_{p(i)}]$,
\begin{equation}
    \xi(s,l) = \xi(t_{i},i) + \int_{t_i}^{s} x(\tilde{s},l) d\tilde{s}.
    \label{eq:eqFlowBetweenPellets}
\end{equation}
with $(s,l) \in \text{dom } q$ for some $l \in \{i,i+1,\ldots,i+n\}$.
In the sequel, we will sometimes write $\xi(s,\cdot)$ for $\xi(s,l)$, in such cases, assuming that $l$ is clear from the context.
Moreover, since for $(t_i,i-1) \in \mathcal{P}(q)$, we have 
$\xi(t_i,i) = 0$ and thus, for all $s \in [t_i,t_{p(i)}]$,
\begin{equation}
    \xi(s,\cdot) = \int_{t_i}^{s} x(\tilde{s},\cdot)d\tilde{s} .
    \label{eq:xi_total_flow}
\end{equation}

Using the jump map in (\ref{eq:jump_map})-(\ref{eq:jump_map_G2}), we have that for each
$k \in \mathbb{Z}_{\geq0}$, with $(t_k, k-1) \in \text{dom }q$ and $(t_k, k) \in \text{dom } q$,
\begin{equation}
    \begin{aligned}
        x(t_k,k) &= x(t_k,k-1) &\text{ when } (t_k,k-1) \notin \mathcal{P}(q) \\
        \hfill x(t_k,k) &= x(t_k,k-1) - \alpha \hfill &\text{ when } (t_k,k-1) \in \mathcal{P}(q)
    \end{aligned}
    \label{eq:reset_x_T2}
\end{equation}

Using (\ref{eq:pellet_jump_time}), we have from (\ref{eq:x_s_i}) that for all $m \in \{0,1,\ldots,n-1\}$ and all $s \in [t_{i+m},t_{i+m+1}]$,
\begin{equation}
    x(s,i+m) = r-e^{-\frac{1}{\tau}(s-t_{i+m})}\left(r-x(t_{i+m},i+m)\right).
\end{equation} 
For all $s \in [t_{i+n},t_{p(i)}]$,
\begin{equation}
    x(s,i+n) = r-e^{-\frac{1}{\tau}(s-t_{i+n})}\left(r-x(t_{i+n},i+n)\right).
\end{equation}
So, from (\ref{eq:reset_x_T2}), we obtain for all $s \in [t_i,t_{p(i)}]$,
\begin{equation}
    x(s,\cdot) = r-e^{-\frac{1}{\tau}(s-t_i)}\left(r-x(t_i,i)\right).
    \label{eq:x_flow_between_pellets}
\end{equation}
From (\ref{eq:pellet_jump_time}) and (\ref{eq:x_flow_between_pellets}), we get
\begin{equation}
    x(t_{p(i)},p_i) = r-e^{-\frac{1}{\tau}(t_{p(i)}-t_i)}\left(r-x(t_i,i)\right),
    \label{eq:V_increase_between_pellets}
\end{equation}
which is the increase in $x$ between two jumps, where a pellet is fired, due to flow (without the effect of pellet injection).

Combining (\ref{eq:xi_total_flow}) with (\ref{eq:x_flow_between_pellets}), we get for all $s \in [t_i,t_{p(i)}]$,
\begin{equation}
\begin{aligned}
    \xi(s,\cdot) &= \int_{t_i}^{s} \left(e^{-\frac{1}{\tau}(\tilde{s}-t_i)}\left(x(t_i,i)-r\right)+r\right)d\tilde{s} \\
     &= \tau (x(t_i,i)-r) \left(1-e^{-\frac{1}{\tau}(s-t_i)}\right)+r(s-t_i).
     \label{eq:neuron_between_pellets}
\end{aligned}
\end{equation}

\noindent \textbf{Computing maximum time between two pellets launches}

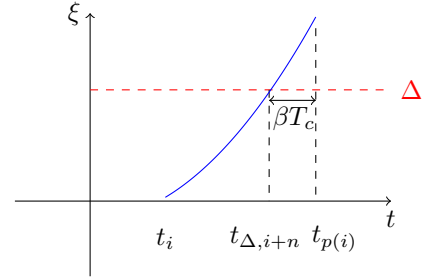
\begin{figure}[thpb]
    \centering
    \begin{tikzpicture}
      \draw[->] (-1, 0) -- (4, 0) node[below] {$t$};
      \draw[->] (0, -1) -- (0, 2.5) node[left] {$\xi$};
      \draw[<->] (2.38,1.34) -- (3, 1.34) node[above] {};
      \draw[dashed,red] (0, 1.48) -- (4, 1.48) node[right] {$\Delta$};
      \draw[dashed] (2.38, 1.48) -- (2.38, 0) node[right] {};
      \draw[dashed] (3, 2.3) -- (3, 0) node[right] {};
      \node (tmp1) at (1,-0.5) {$t_i$};
      \node (tmp2) at (2.33,-0.5) {$t_{\Delta, i+n}$};
      \node (tmp3) at (3.25,-0.5) {$t_{p(i)}$};
      \node (tmp4) at (2.71,1.1) {$\beta T_c$};
      \draw[domain=1:3, smooth, variable=\x, blue] plot ({\x}, {0.3*\x*\x-0.25});
    \end{tikzpicture}
    \caption{At time $t_i$, $\xi=0$, when $\xi$ reaches the threshold $\Delta$ at time $t_{\Delta, i+n}$, it needs to wait for the timer $T$ to reach $T_c$ before triggering a jump at $t_{p(i)}$.}
    \label{fig:xi_overshoot}
   \end{figure}

To capture the effect of synchronising the neuron spike to the timer, we define $t_{\Delta, i+n} := \inf\{s\geq t_{i+n} : \xi(s, i+n) = \Delta\}$ as the time the neuron membrane potential $\xi$ reaches its threshold $\Delta$.
Jump set $\mathcal{D}_2$ in (\ref{eq:jumpsetD2}) shows that when $\xi$ reaches $\Delta$, the jump will occur at most $T_c$ time units later than $t_{\Delta, i+n}$,
which we write as $\beta T_c$, for some $0\leq \beta < 1$, so that $t_{p(i)}-t_i = k_c T_c$, with $k_c \in \mathbb{Z}_{>0}$. A visual representation is given in Fig. \ref{fig:xi_overshoot}.

Thus, at $t_{\Delta, i+n}$, (\ref{eq:neuron_between_pellets}) becomes
\begin{equation}
\begin{aligned}
    \xi(&t_{\Delta, i+n},i+n) \\
    &= r(t_{\Delta, i+n}-t_i) - \tau (r-x(t_i,i)) \left(1-e^{-\frac{1}{\tau}(t_{\Delta, i+n}-t_i)}\right) \\ 
    &= \Delta .
    \label{eq:xi_at_t_delta}
\end{aligned}
\end{equation}
Using (\ref{eq:pellet_jump_time}) and (\ref{eq:xi_at_t_delta}), and since $t_{\Delta, i+n} = t_{p(i)}-\beta T_c$, we get for each
$i \in \{0,1,\ldots,j\}$ such that $(t_i,i-1) \in \mathcal{P}(q)$,
\begin{multline}
t_{p(i)}-t_i = \\
\frac{1}{r}\left(\Delta+\tau
    (r-x(t_i,i)) \left(1-e^{-\frac{1}{\tau}(t_{p(i)}-t_i-\beta T_c)}\right) \right)+\beta T_c.
    \label{eq:dwell_time}
\end{multline}
Using $x(t_i,i) \in (0,r]$ and $\beta \in [0,1)$, we get
\begin{equation}
t_{p(i)}-t_i \leq \frac{\Delta}{r}+\tau
     \left(1-e^{-\frac{1}{\tau}(t_{p(i)}-t_i-T_c)}\right) + T_c .
\end{equation}
Given that
$\Delta \leq r \tau \ln{\frac{r}{r-\alpha}} - r \tau \left(1- \frac{r-\alpha}{r} e^{\frac{T_c}{\tau}}\right) - r T_c$, by the hypothesis of Theorem~\ref{theorem1},
this becomes
\begin{equation}
t_{p(i)}-t_i \leq \tau \ln{\frac{r}{r-\alpha}} + \tau \frac{r-\alpha}{r} e^{\frac{T_c}{\tau}}-\tau
     e^{-\frac{1}{\tau}(t_{p(i)}-t_i)} e^{\frac{T_c}{\tau}} .
\end{equation}
Rearranging the terms, gives 
\begin{equation}
t_{p(i)}-t_i +\tau e^{-\frac{1}{\tau}(t_{p(i)}-t_i)} e^{\frac{T_c}{\tau}} \leq \tau \ln{\frac{r}{r-\alpha}} + \tau \frac{r-\alpha}{r} e^{\frac{T_c}{\tau}}.
\label{eq:dwell_time_constant_RHS}
\end{equation}

\begin{figure}[thpb]
    \centering
    \includegraphics[width=0.9\linewidth]{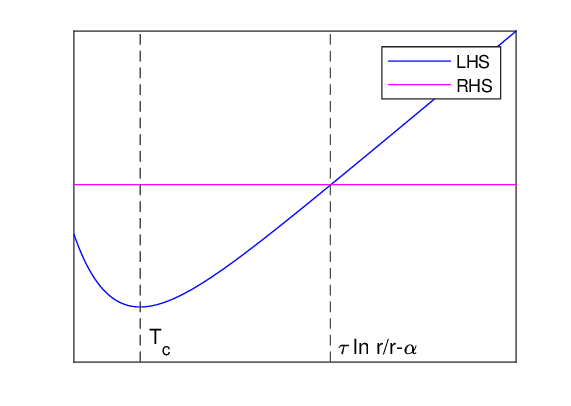}
    \caption{LHS and RHS of (\ref{eq:dwell_time_constant_RHS}) in function of the time $\tau_{d}^i$}
    \label{fig:dwell_time_shape}
\end{figure}

To find the maximum time $t_{p(i)} - t_i$ between two pellet launches, we evaluate the left-hand side (LHS) as a function of the time $\tau_{d}^i:=t_{p(i)}-t_i$, and compare it to the right-hand side (RHS) of (\ref{eq:dwell_time_constant_RHS}), which is constant.
The shape of the LHS compared to a constant RHS is illustrated in Fig. \ref{fig:dwell_time_shape}.
Recall that the minimum dwell-time is determined by the actuator (related to timer $T$ in the model), so  at least it holds that
\begin{equation}
   \tau_{d}^i \geq T_c.
\end{equation}
Taking the derivative of the LHS of (\ref{eq:dwell_time_constant_RHS}), we get
\begin{equation}
    \frac{d}{d\tau_{d}^i} \left(\tau_{d}^i +\tau e^{-\frac{1}{\tau}\tau_{d}^i} e^{\frac{T_c}{\tau}} \right) 
    =1-e^{\frac{T_c}{\tau}}e^{-\frac{\tau_{d}^i}{\tau}}.
    \label{eq:dwelltime_derivative}
\end{equation}
We can verify that (\ref{eq:dwelltime_derivative}) has a zero at $\tau_{d}^i = T_c$, and is positive for all $\tau_{d}^i \geq T_c$, so the LHS of (\ref{eq:dwell_time_constant_RHS}) is monotonically increasing for $\tau_{d}^i \geq T_c$. 
Also, since the equality in (\ref{eq:dwell_time_constant_RHS}) holds at $\tau_{d}^i = \tau \ln{\frac{r}{r-\alpha}}$, the \emph{maximum} time
between two pellets for which (\ref{eq:dwell_time_constant_RHS}) holds when $x(t_i,i-1)>0$, can be simplified to
\begin{equation}
t_{p(i)}-t_i = \tau_{d}^i \leq \tau \ln \frac{r}{r-\alpha}=:\tau_d.
\label{eq:max_dwell_time}
\end{equation}

\noindent \textbf{Decrease of $x$ during pellet launch}

We now consider the decrease in $x$ during a jump due to a pellet launch in $x(t_{p(i)}, i+n) = x(t_{p(i)}, p_i -1) \in \mathcal{D}_2$.
From (\ref{eq:jump_x}) and (\ref{eq:pellet_jump_time}), we have,
\begin{equation}
    x(t_{p(i)},p(i)) = x(t_{p(i)},i+n)-\alpha.
    \label{eq:V_decrease_pellet}
\end{equation}

\noindent \textbf{Guaranteeing decrease of flow and jump between pellet launches}

Given the increase in $x$ in (\ref{eq:V_increase_between_pellets}), the maximum time between two consecutive pellet jumps in (\ref{eq:max_dwell_time}), and the decrease in $x$ in (\ref{eq:V_decrease_pellet}),
we now show that there is a net decrease of $x$ over one cycle, consisting of increase during flow, and decrease in one pellet launch. 
Combining (\ref{eq:V_decrease_pellet}) with (\ref{eq:V_increase_between_pellets}), we get
\begin{equation}
    x(t_{p(i)},p(i)) = r-e^{-\frac{1}{\tau}(t_{p(i)}-t_i)}\left(r-x(t_i,i)\right) -\alpha ,
    \label{eq:x_flow_and_pelletjump}
\end{equation}
where we recall that the cycle starts right after a previous pellet launch at $(t_i,i)$, with $x(t_i,i) >0$ the error at the beginning of the cycle, and $x(t_{p(i)},p(i))$ the error at the end, after the next pellet launch. 
Since $x(t_i,i) \in (0,r]$, and using (\ref{eq:max_dwell_time}), this becomes
\begin{equation}
\begin{split}
    x(t_{p(i)},p(i)) 
    =& \Bigg(\frac{r\left(1-e^{-\frac{1}{\tau}(t_{p(i)}-t_i)}\right)-\alpha}{x(t_i,i)}
    \\&+e^{-\frac{1}{\tau}(t_{p(i)}-t_i)}\Bigg)
    x(t_i,i) \\
    \leq& \left(\frac{r\left(1-\frac{r-\alpha}{r}\right)-\alpha}{x(t_i,i)}
    +\frac{r-\alpha}{r}\right) x(t_i,i) \\
    =& \; \frac{r-\alpha}{r} \; x(t_i,i) \\
    =& \; \gamma x(t_i,i) ,
\end{split}
\label{eq:x_decrease_flow_and_jump}
\end{equation}
with $\gamma = \frac{r-\alpha}{r} < 1$.
Therefore,
for all $i \in \{1,2\ldots,j\}$ such that $(t_i,i-1) \in \mathcal{P}(q)$ and $x(t_i, i-1) >0$, 
\begin{equation}
    x(t_{p(i)},p(i)) \leq \gamma x(t_i,i) ,
    \label{eq:jumpDecrease}
\end{equation}
meaning that after a full cycle, including a flow starting immediately after the last pellet jump in $(t_i,i)$, and including the pellet jump in $(t_{p(i)}, p(i))$, $x$ has net decreased.
Using the jump map in (\ref{eq:jump_map})-(\ref{eq:jump_map_G2}), and recalling that the jump size is equal to $\alpha$, we have that for all 
$s \in [t_i,t_{p(i)}]$ such that $x(t_i,i) \in (0,r]$,
\begin{equation}
    -\alpha < x(s,\cdot) \leq \gamma x(t_i,i) + \alpha .
    \label{eq:stabilityOneFlowOneJump}
\end{equation}
Pick any $(t^\star, j^\star-1), (t,j) \in \text{dom } q$ such that  $(t^\star, j^\star-1) \in \mathcal{P}(q) \cup (0,0)$,  $(t^\star,j^\star-1) \leq (t,j)$ and $x(t^\star, j^\star-1)>0$. Denote $n(t,t^\star)$ the number of pellet launches that occur between $(t^\star,j^\star-1)$ and $(t,j)$, in view of \eqref{eq:max_dwell_time} we have that 
\begin{equation}
    t-t^\star \leq (n(t,t^\star) +1) \tau_d, 
    \label{eq:MaxDwellTimeTot}
\end{equation}
which corresponds to 
\begin{equation}
    n(t,t^\star) \geq \frac{t-t^\star}{\tau_d} -1. 
    \label{eq:maximumAverageDwellTime}
\end{equation}
Moreover, using \eqref{eq:stabilityOneFlowOneJump} we have that, for all $s \in (t^\star,t)$,
\begin{equation}
    -\alpha < x(s,\cdot) \leq \gamma^{n(t,t^\star)} x(t^\star,j^\star) + \alpha ,
    \label{eq:stabilityOneFlowOneJump2}
\end{equation}
which, since $\gamma \in (0,1)$ and using \eqref{eq:maximumAverageDwellTime}, becomes
\begin{equation}
    -\alpha < x(s,\cdot) \leq \gamma^{\frac{t-t^\star}{\tau_d} -1} x(t^\star,j^\star) + \alpha ,
\end{equation}
Consequently, we have that for all 
$(t,j) \in \text{dom }q$ such that $x(0,0)>0$
\begin{equation}
    -\alpha < x(t,j) \leq \gamma^{\frac{t}{\tau_d} -1} x(0,0) + \alpha .
    \label{eq:xPosFinal}
\end{equation}
Moreover, since the solution $q$ is $t$-complete (see Remark \ref{rem:completeness}), from \eqref{eq:xPosFinal} we have that, for any $x(0,0) >0$, any corresponding maximal solution to \eqref{eq:hybrid_system} satisfies 
$\lim_{t+j \rightarrow \infty} \sup |x(t,j)| \leq \alpha$. 

Now consider the case where $x(t_i,i) \leq 0$.
From the neuron dynamics in (\ref{eq:neuron_flow}), it is evident that no control actions occur when $x \leq 0$, as $\max{(0,x)} = 0$ in \eqref{eq:neuron_flow}. 
Since (\ref{eq:openloop_x}) shows that $x$ will always increase without control actions the controller will ``wait'' until it starts working again, starting from $(\bar t, i) \in \text{dom } q$ such that $x(\bar t,i)=0$ and $x(\bar s,i) \geq 0$ for all $\bar s \in (\bar t,t_{p_i}]$. Thus, for all $\bar s \in (\bar t,t_{p_i}]$, following similar steps as in the case where $x(t_i,i)> 0$, we have
\begin{equation}
    -\alpha < x(\bar s,i) \leq \alpha. 
    \label{eq:xNeg1}
\end{equation}
Moreover, for all $s \in [t_i,\bar t]$, 
 from (\ref{eq:openloop_x}), we have 
\begin{equation}
    x(s,i) = r-e^{-\frac{s-t_i}{\tau}}\left(r-x(t_i,i)\right).
    \label{eq:xNegFinal1}
\end{equation} 
Consequently, from \eqref{eq:xNeg1} and \eqref{eq:xNegFinal1}, we have that for all $(t,j) \in \text{dom }(q)$, such that with $x(0,0) \leq 0$:
\begin{align}
   \text{min}\left(r-e^{-\frac{t}{\tau}}\left(r-x(0,0)\right),-\alpha\right) < x(t,j) \leq  \alpha .
   \label{eq:xNegFinal2}
\end{align}
Moreover, merging \eqref{eq:xPosFinal} and \eqref{eq:xNegFinal2} and since the solution $q$ is $t$-complete (see Remark \ref{rem:completeness}), we have that, for any $x(0,0) \in \R$, any corresponding maximal solution to \eqref{eq:hybrid_system} satisfies 
$\lim_{t+j \rightarrow \infty} \sup |x(t,j)| \leq \alpha$. 
This concludes the proof.\hfill $\blacksquare$


\end{document}